\documentclass[conference]{IEEEtran}
\usepackage{amsmath,amssymb,amsfonts}
\usepackage{algorithmic}
\usepackage{graphicx}
\usepackage{textcomp}
\usepackage{xcolor}
\usepackage{url}
\usepackage{hyperref}

\newboolean{showcomments}
\setboolean{showcomments}{true}

\newboolean{showboxes}
\setboolean{showboxes}{true}

\usepackage[utf8]{inputenc}
\usepackage{url}
\usepackage{colortbl}
\usepackage{balance}
\usepackage{xspace}
\usepackage{graphicx}
\usepackage{verbatim}
\usepackage{bold-extra}
\usepackage{paralist}
\usepackage{multirow}
\usepackage{listings}
\usepackage{boxedminipage}
\usepackage{soul}
\usepackage{enumitem}
\usepackage[normalem]{ulem}
\setlist{nolistsep,leftmargin=.5cm}
\usepackage{booktabs}
\usepackage{tabularx}
\usepackage{svg}
\usepackage{calc}
\usepackage{float}
\usepackage{multirow}
\usepackage[normalem]{ulem}
\useunder{\uline}{\ul}{}
\usepackage[most]{tcolorbox}
\usepackage{caption}
\usepackage{microtype} 
\usepackage[noadjust]{cite}
\usepackage{graphicx}
\usepackage{textcomp}
\usepackage{xcolor}
\usepackage{hyperref}
\usepackage{ifthen}
\usepackage{wrapfig}
\usepackage{subcaption}
\usepackage{cleveref} 
\usepackage{adjustbox}
\usepackage{totcount}

\usepackage{breakurl}

\Crefname{figure}{Fig.}{\textbf{Figs.}}
\Crefname{table}{Tab.}{\textbf{Tabs.}}
\Crefname{section}{Sec.}{\textbf{Secs.}}

\ifthenelse{\boolean{showcomments}}
{\newcommand{\nb}[2]{
		\fbox{\bfseries\sffamily\scriptsize#1}
		{\sf\small$\blacktriangleright$\textit{#2}$\blacktriangleleft$}
	}
	
}
{\newcommand{\nb}[2]{}
	
}

\newcommand\rev[1]{{\color{black}{#1}}}

\newcommand{\ie}{\textit{i.e.},\xspace}
\newcommand{\eg}{\textit{e.g.},\xspace}
\newcommand{\etc}{\textit{etc.}\xspace}
\newcommand{\etal}{\textit{et al.}\xspace}

\newcommand{\hqc}{HQC\xspace}
\newcommand{\xan}{XDF\xspace}
\newcommand{\stackexchange}{QCSE\xspace}
\newcommand{\quantumcomputing}{QC\xspace}

\newcounter{findingcounter}
\newtotcounter{totalfindings}
\ifthenelse{\boolean{showboxes}}
{
    \newcommand{\finding}[1]{%
      \refstepcounter{findingcounter}
      \begin{tcolorbox}[boxsep=1pt,left=2pt,right=2pt,top=1pt,bottom=1pt]%
      \small
      \centering
      \textbf{Finding \arabic{findingcounter}:} #1
      \end{tcolorbox}%
      \addtocounter{totalfindings}{1}
    }
    
  }
{
    \newcommand{\finding}[1]{}
}

\begin{document}

\title{When Quantum Meets Classical: Characterizing Hybrid Quantum-Classical Issues Discussed in Developer Forums}

\author{
\IEEEauthorblockN{Jake Zappin}
\IEEEauthorblockA{
\textit{William \& Mary}\\
Williamsburg, Virginia, USA \\
azappin@wm.edu}
\and
\IEEEauthorblockN{Trevor Stalnaker}
\IEEEauthorblockA{\textit{William \& Mary}\\
Williamsburg, Virginia, USA \\
twstalnaker@wm.edu}
\and
\IEEEauthorblockN{Oscar Chaparro}
\IEEEauthorblockA{
\textit{William \& Mary}\\
Williamsburg, Virginia, USA \\
oscarch@wm.edu}
\and
\IEEEauthorblockN{Denys Poshyvanyk}
\IEEEauthorblockA{\textit{William \& Mary}\\
Williamsburg, Virginia, USA \\
denys@cs.wm.edu}
}
\vspace{-2cm}
\maketitle

\begin{abstract}
Recent advances in quantum computing have sparked excitement that this new computing paradigm could solve previously intractable problems.  However, due to the faulty nature of current quantum hardware and quantum-intrinsic noise, the full potential of quantum computing is still years away.  Hybrid quantum-classical computing has emerged as a possible compromise that achieves the best of both worlds.  In this paper, we look at hybrid quantum-classical computing from a software engineering perspective and present the first empirical study focused on characterizing and evaluating recurrent issues faced by developers of hybrid quantum-classical applications. The study comprised a thorough analysis of 531 real-world issues faced by developers -- including software faults, hardware failures, quantum library errors, and developer mistakes -- documented in discussion threads from forums dedicated to quantum computing.  By qualitatively analyzing such forum threads, we derive a comprehensive taxonomy of recurring issues in hybrid quantum-classical applications that can be used by both application and platform developers to improve the reliability of hybrid applications. The study considered how these recurring issues manifest and their causes, determining that hybrid applications are crash-dominant (74\% of studied issues) and that errors were predominantly introduced by application developers (70\% of issues). We conclude by identifying recurring obstacles for developers of hybrid applications and actionable recommendations to overcome them.
\looseness=-1
\end{abstract}

\section{Introduction}

In the last decade, quantum computing (\quantumcomputing) has transitioned from a theoretical concept to a practical endeavor \cite{preskill2023quantum}.  This has been primarily fueled by the limitations of classical computing hardware \cite{nagy2007quantum} and the emergence of accessible \quantumcomputing platforms like IBM's Qiskit \cite{aleksandrowicz2019qiskit}, Google's Cirq \cite{cirqmain}, Microsoft’s Q\# \cite{svore2018q} and Xanadu’s PennyLane \cite{bergholm2018pennylane}.  \quantumcomputing is not merely an incremental step in computing power, but an entirely new kind of computing paradigm that has the potential to fundamentally change the way we represent information and perform computations \cite{bhasin2021quantum}. Indeed, some computations that take years on modern classical computers may be solved in minutes by quantum computers due to the leap in computational power \cite{jozsa2003role}. 

However, \quantumcomputing is still in its infancy.  For instance, although significant progress has been made, it is unlikely that fault-tolerant quantum computers will be developed in the near-term~\cite{battistel2023real}.  As a result, many developers and researchers have turned to \textit{hybrid quantum-classical} (\hqc) computing as a bridge to solve complex problems that may be out of reach for classical computers alone~\cite{callison2022hybrid}. This approach leverages the respective capabilities of quantum and classical systems. It enables quantum devices to efficiently handle specific computational tasks within an algorithm that classical computers find intractable, while classical systems manage the remaining tasks of the process. This convergence of technologies can enhance an algorithm’s overall performance 
over a purely classical algorithm designed for the same task, particularly for applications like machine learning ~\cite{arthur2022hybrid}, even on today’s Noisy Intermediate-Scale Quantum (NISQ)-era computers \cite{endo2021hybrid, houssein2022hybrid}.

The promise of these performance gains has recently prompted developers and researchers to think about \quantumcomputing from a software engineering (SE) perspective \cite{piattini2021toward, ali2022software, zhao2020quantum, barbosa2020software}.  This includes the consideration of essential SE tasks,\ such as debugging and testing, in the context of \quantumcomputing \cite{mendiluze2021muskit, paltenghi2022bugs}.  However, developing quantum applications can pose unique challenges due to the complexity of quantum mechanics, the introduction of quantum principles like superposition and entanglement \cite{li2022bug} and the relative infancy of \quantumcomputing platforms.  Quantum developers therefore face SE issues and challenges that are new and unique, and have not yet been fully studied.

Hence, it is crucial to develop methodologies and taxonomies for identifying, classifying, and fixing issues faced by developers working on systems incorporating quantum algorithms.  To date, though, attention has been mainly focused on identifying bugs \textit{in} quantum platforms and libraries~\cite{paltenghi2022bugs, zhao2021identifying, luo2022comprehensive, zhao2021bugs4q, zhao2023empirical} as opposed to exploring recurring issues, including recurrent bug patterns, encountered by developers of quantum algorithms and applications, the end-users of those platforms/libraries.  More importantly, there have been no studies specifically looking at recurring issues in \hqc systems.

\hqc algorithms and applications, therefore, represent a particularly essential ground for research in SE  due to their potential to overcome performance challenges on purely classical hardware.  Studying the recurring issues encountered by developers within these hybrid systems is critical for several reasons: 
\looseness=-1
\begin{enumerate}

    \item \textit{Prevalence of \hqc systems.} Due to their practicality, \hqc systems make up most of the current quantum applications being developed (variational quantum eigensolvers, quantum approximate optimization algorithms, hybrid quantum-classical machine learning  models, \etc~\cite{mcclean2016theory}).  This makes them the most relevant for immediate industry and research utilization \cite{callison2022hybrid}; in fact, \hqc algorithms and applications are already impacting industries such as finance, material science, logistics, and healthcare \cite{houssein2022hybrid, fernandez2021hybrid, ding2021implementation, rosmanis2022hybrid}.
    
    \item \textit{A novel programming paradigm.} \quantumcomputing represents a new programming paradigm and requires a fundamentally different set of skills and problem-solving techniques.  Also, the interaction between classical and quantum parts introduces unique challenges that are absent in purely classical systems — challenges that must be understood and mitigated to realize the potential of quantum-enhanced computing \cite{callison2022hybrid}.
    
    \item \textit{Inform research and development.} The study of recurring \hqc system issues encountered by developers will not only prepare us for the arrival of more advanced quantum systems, but also inform the development of current quantum algorithms, libraries, and platforms.  These insights provide immediate benefits to a broad range of stakeholders (developers, researchers, industry, government, \etc) who are at the forefront of quantum development \cite{walsh2023past, bayerstadler2021industry}.
\end{enumerate}

Accordingly, we present the first empirical study to characterize and evaluate recurring issues faced by developers in \hqc algorithms, applications, and systems (hereafter referred to collectively as \textit{applications}).  We collected and inspected \textit{531} real-world issues encountered by developers and researchers reported on two different sources: the Xanadu Discussion Forums (\xan) \cite{xanaduforums} (now known as the PennyLane Discussion Forums) and the Quantum Computing Stack Exchange (\stackexchange) message board \cite{quantumstackexchange}.  By employing a rigorous methodology to collect and analyze recurring issues in hybrid systems, we aim to answer five research questions, which includes developing a comprehensive taxonomy of recurring issues encountered by developers in \hqc applications. 

The findings of this study, presented \Cref{sec:results}, offer valuable insights into the current issues and challenges facing \hqc developers, as they aim to guide researchers and developers in formulating effective strategies for mitigating them. To  facilitate more research in this domain, we make publicly available our \hqc issue dataset, annotations, and results~\cite{anonymous_repo}.
\looseness=-1

In summary, this paper makes the following contributions:

\begin{itemize}

\item A novel comprehensive study of recurring real-world issues faced by developers of \hqc algorithms and applications;

\item An in-depth review of 531 real-world issues in \hqc implementations, documented in 447 discussion threads; 

\item A thorough taxonomy of recurring \hqc issues \rev{intended to assist both platform and application developers in troubleshooting issues and in prioritizing 
the development of testing and debugging tools};

\item A comprehensive analysis of the taxonomy, including quantitative and qualitative analyses of the manifestations and causes of the issues to assist developers in implementing \hqc applications and in improving the usability and robustness of quantum platforms and libraries;

\item A labeled dataset of real-world recurring issues faced by developers that can enable future research \cite{anonymous_repo}.

\end{itemize}

\section{Background}

\quantumcomputing represents a significant paradigm shift in CS and, more specifically, in SE~\cite{hevia2021quantum}.  It is a multidisciplinary area of computing that harnesses the principles of quantum mechanics to process information in ways fundamentally different from classical computing\rev{~\cite{zhao2020quantum, barbosa2020software}}).  This section briefly introduces \quantumcomputing, its challenges, and the concept of \hqc computing.

\subsection{The Qubit, Superposition, and Entanglement}

Qubits are the fundamental units of \quantumcomputing, differing from classical bits by utilizing quantum superposition and entanglement~\cite{hey1999quantum}. Superposition permits a qubit to represent multiple states concurrently.  This permits quantum computers to process numerous calculations simultaneously offering a potential exponential speed-up for certain computations over classical computers \cite{rieffel2000introduction}. The phenomenon of entanglement creates a dependency between the states of qubits, such that an action on one qubit can affect another instantaneously, a phenomenon Einstein called ``spooky action at a distance" \cite{einstein1971born}. These quantum characteristics are what give quantum computers the ability to perform complex operations in tandem and potentially solve problems that are currently unsolvable by classical systems \cite{rieffel2000introduction, mcardle2020quantum, shaydulin2023evidence}.

The utilization of superposition and entanglement grants \quantumcomputing the capacity to outperform classical systems in specific areas. Such capabilities enable the crafting of quantum algorithms that substantially lower time complexity for complex problems and address tasks deemed intractable for classical computers due to computational limitations \cite{mcardle2020quantum, shaydulin2023evidence}. A notable example is prime number factorization, essential for modern encryption, where \quantumcomputing, through Shor’s algorithm, achieves polynomial-time factoring, a significant reduction from classical methods' exponential time~\cite{bhatia2020efficient}. As \quantumcomputing technology continues to advance, it is anticipated that we will not only develop more efficient quantum algorithms to replace many classical algorithms, but also solve problems currently beyond the reach of classical computing~\cite{perdomo2018opportunities}.
\looseness=-1

\subsection{\quantumcomputing Limitations and the NISQ-Era}

Despite its potential, quantum computing faces developmental hurdles due to qubit instability, caused by environmental interference and other intrinsic factors leading to errors and decoherence \cite{preskill2018quantum}. These issues necessitate error-correcting qubits, adding complexity and the risk of additional errors \cite{preskill1998fault}. Achieving fault tolerance is anticipated to require a quantum processor with millions of qubits \cite{gottesman2010introduction}, a stark contrast to the Noisy Intermediate-Scale Quantum (NISQ) devices of today which have up to one thousand qubits \cite{ibm1000qubits, ibm400qubits}. 

 This current NISQ-era is characterized by quantum computers with 10 to 1,000+ noisy qubits and no error correction~\cite{bharti2022noisy}. These systems use probabilistic methods, running quantum programs multiple times to obtain likely outcomes \cite{tannu2019not}. Despite limitations, NISQ-era quantum computers offer advancements, exemplified by algorithms like the Variational Quantum Eigensolver (VQE), which perform beyond classical capabilities with respect to specific tasks \cite{bharti2022noisy}. NISQ-era machines are consequently instrumental in advancing (\hqc) computing architectures, as they are being employed to augment and enhance today's classical computing systems.

\subsection{Bridging the Gap: \hqc Computing}
\label{sec:bridge}

Hybrid quantum-classical (\hqc) systems are an intermediate step towards future fault-tolerant quantum computers, merging quantum computational power with classical reliability and accessibility \cite{endo2021hybrid}. These systems utilize quantum computers for complex tasks, such as prime factorization, optimization and other intractable problems for current classical computers, while relying on classical computing for standard operations and processing, thereby avoiding the current quantum limitations like qubit instability and error correction challenges \cite{endo2021hybrid, phillipson2023classification}.

The synergy in \hqc architectures enhances performance beyond what quantum or classical systems can achieve alone~\cite{callison2022hybrid, phillipson2023classification}. This approach fosters scalable, real-world quantum applications. Notably, \hqc algorithms like the Variational Quantum Eigensolver (VQE) excel in quantum chemistry, providing high-precision estimates unattainable by classical systems \cite{phillipson2023classification, chan2023hybrid}, and are adaptable across various disciplines. The Quantum Approximate Optimization Algorithm (QAOA) is another \hqc algorithm that addresses combinatorial problems by combining quantum and classical techniques for simultaneous solution exploration and error correction \cite{phillipson2023classification, acampora2023genetic}. \hqc is also advancing machine learning (ML), an area where the limitations of classical computers are particularly being felt \cite{de2022survey, wossnig2021quantum}.
\looseness=-1

Consequently, \hqc computing is not just a workaround for the current limitations of quantum technology.  Rather, it is a strategic approach that combines the best of both the quantum and classical worlds to increase computing power and efficiency.  Given that \hqc algorithms and applications will become more and more important in the coming years, it is critical to study them from a SE perspective. \rev{This is particularly so given that \hqc introduces new challenges like requiring seamless integration of quantum and classical components, including managing data exchange, embedding, encoding/decoding, and synchronization. Additionally, \hqc involves balancing quantum hardware limitations with classical capabilities for resource management and performance optimization. These complexities demand specialized expertise and tailored approaches, making \hqc programming a distinct paradigm in its own right requiring dedicated research.}

\section{Related Work}

The introduction and rise of \quantumcomputing has introduced novel challenges from a SE perspective. Recent studies have focused on investigating bug patterns in quantum libraries~\cite{zhao2023empirical}, platforms~\cite{paltenghi2022bugs}, and languages \cite{luo2022comprehensive}, with a focus on quantum-specific bugs and automated detection \cite{nayak2023q}. Luo \etal identified 80\% of bugs as quantum-specific in various quantum programming languages \cite{luo2022comprehensive}, while Paltenghi and Pradel found 39.9\% of platform bugs were unique to quantum platforms \cite{paltenghi2022bugs}.

Zhao \etal's work includes the development of the Bugs4Q suite \cite{zhao2021bugs4q} and tests for certain recurring bugs in IBM Qiskit platform, despite its limitations due to recent deprecations in Qiskit~\cite{qiskitdeprecation}.  Zhao \etal has more recently studied bugs in quantum ML frameworks developing a taxonomy of recurring bug patterns \cite{zhao2023empirical}. Similarly, Nayak \etal's Q-PAC framework marks progress in bug-fix pattern detection in quantum code~\cite{nayak2023q}.

Aoun \etal~\cite{li2022bug} performed an analysis of quantum GitHub repositories identifying prevalent bug types in quantum software projects, which included simulators, frameworks (\eg Qiskit, Cirq), algorithms, compilers, tools, and experimental projects.  The study revealed quantum systems to be generally buggier than classical ones, with program anomalies, configuration issues, and data structure flaws as common problems.
\looseness=-1

Our study complements and extends prior research into quantum bugs in several ways.  First, while our study encompasses bug identification, the scope of our study is broader.  Specifically, it attempts to understand and categorize recurring issues and challenges faced by quantum developers, particularly given \quantumcomputing and \hqc are in their infancy. By doing so, we hope to inform the development of methodologies and the evolution of quantum libraries and frameworks still in their infancy.
Second, we examine the relatively unexplored domain of issues (including recurring bugs) in \hqc systems. We focus on hybrid algorithms and applications as a whole rather than solely on quantum components. \rev{Past studies have separated classical and quantum bugs without considering these hybrid applications or the interaction between the two subsystems. By focusing on hybrid applications, we identified a new type of issue: \textit{cross-domain issues}.}

Third, our research methodology also differs by sourcing data from \xan and \stackexchange, providing community-driven insights into the obstacles faced by developers in this emerging field~\cite{xanaduforums, quantumstackexchange}. \rev{As \Cref{tab:comparison} illustrates, prior research has focused on mining data primarily from GitHub repositories of quantum platforms, frameworks, and libraries. Our research is fundamentally different in that we used quantum-specific forums, which have been left unexplored. We believe these forums are a far richer resource for mining bugs and issues in quantum software, offering more specialized and relevant insights into the challenges unique to developers of quantum applications.}

Lastly, and most importantly, our study focuses on recurring issues in HQC \textit{applications} and \textit{algorithms}. \rev{Prior research has predominantly focused on identifying bugs in quantum platforms and libraries from the perspective of platform developers (see \Cref{tab:comparison}). These developers are concerned with the infrastructure and tools that enable application development, such as Qiskit, Cirq, and Q\#. In contrast, application developers, often found on the \xan and \stackexchange forums, focus on building quantum applications and algorithms to solve particular problems, like VQE and QAOA, using these platforms and libraries. By examining applications, we provide insights relevant to end-users of these platforms, offering information to improve both application and platform development and maintenance.}

\rev{

\begin{table}[ht]
\caption{\rev{Comparison with prior related work}}
\label{tab:comparison}
\resizebox{\columnwidth}{!}{
\begin{tabular}{c|c|c}
\textbf{Paper} & \textbf{Data Sources} & \textbf{Software Studied} \\ \hline
\begin{tabular}[c]{@{}c@{}}This \\ Paper\end{tabular} & \begin{tabular}[c]{@{}c@{}}Xanadu Discussion Forums (\xan) \\ and QC Stack Exchange (\stackexchange) \end{tabular} & \begin{tabular}[c]{@{}c@{}}Quantum Applications and Algorithms \\ (as opposed to platforms and frameworks)\end{tabular} \\ \hline
\cite{paltenghi2022bugs} & \begin{tabular}[c]{@{}c@{}}GitHub Repositories \\ (\ie Issues and Pull Requests)\end{tabular} & Quantum Platforms (Qiskit, Cirq, Q\#, etc.) \\ \hline
\cite{li2022bug} & \begin{tabular}[c]{@{}c@{}}GitHub Repositories \\ (\ie API Events)\end{tabular} & \begin{tabular}[c]{@{}c@{}}Quantum Simulators, Platforms, \\ Compilers, Tools, and Libraries\end{tabular} \\ \hline
\cite{luo2022comprehensive} & \begin{tabular}[c]{@{}c@{}}(1) GitHub Repositories, \\ (2) Stack Overflow Posts; and \\ (3) Stack Exchange Posts\end{tabular} & \begin{tabular}[c]{@{}c@{}}Quantum Platforms (Qiskit, Cirq, \\ Q\# and ProjectQ)\end{tabular} \\ \hline
\cite{zhao2023empirical} & \begin{tabular}[c]{@{}c@{}}GitHub Repositories \\ (\ie Issue Reports)\end{tabular} & \begin{tabular}[c]{@{}c@{}}Quantum ML Platforms (Qiskit, Q\#, \\ Cirq, Torch Quantum, PennyLane)\end{tabular} \\ \hline
\cite{zhao2021bugs4q} & GitHub Repositories & Quantum Platforms (Qiskit) \\ \hline
\cite{nayak2023q} & \begin{tabular}[c]{@{}c@{}}(1) Prior Research, \\ (2) Stack Exchange Posts, and \\ (3) GitHub Repositories\end{tabular} & Quantum Platforms (Qiskit) \\ % Blank row
\end{tabular}%
}
\end{table}

}

\section{Study Design}

This study aims to characterize recurring issues encountered by developers of \hqc applications, their manifestations, and their causes, as well as to identify the major challenges that practitioners face when developing \hqc applications. Our intention is to offer actionable insights to practitioners that allow them to develop better and more streamlined \hqc applications. To this end, we reviewed the most comprehensive quantum-specific sources available: \xan and \stackexchange, where developers, students, and researchers (\textit{developers} from hereon) discuss issues encountered while writing \hqc applications.

This study answers the following research questions (RQs):

\textbf{RQ1:} \textit{What are the recurring issues encountered by developers of \hqc applications?} This RQ aims at identifing recurring issues that arise when developing \hqc algorithms and applications, providing a foundation for developing strategies to detect and prevent these problems.  The data collected in answering this RQ was used to synthesize the taxonomy in \Cref{fig:taxonomy}.

\textbf{RQ2:} \textit{How frequently are issues encountered in \hqc applications quantum-specific, classical, or cross-domain?}
This RQ aims to categorize issues based on their origin and the interaction between quantum and classical components. We hope this offers insights into the unique operational challenges of \hqc systems and where issues typically manifest, which can guide more targeted resolution efforts and enhance overall application efficiency and reliability.

\textbf{RQ3:} \textit{How do issues in \hqc applications manifest?} \hqc applications can be inherently more complex than purely classical ones due to the intertwining of non-intuitive quantum behaviors with traditional classical computing logic.  This RQ delves into ways  \hqc issues appear and manifest to developers, considering the unique challenges posed by the integration of quantum and classical paradigms.

\textbf{RQ4}: \textit{What are the root causes of issues encountered in \hqc applications?}  
This RQ examines whether recurring issues stem from the complexities of quantum programming, platform updates, documentation gaps, or other factors.

\textbf{RQ5:} \textit{What are the predominant challenges and issues faced by developers when developing \hqc applications?} There are unique obstacles presented by \quantumcomputing, the rapidly evolving quantum ecosystem of platforms and libraries, and the inherent complexity of merging with classical computing systems. This RQ attempts to identify the predominant challenges developers encounter while writing \hqc applications and offers suggestions on how these challenges may be mitigated.

\subsection{Data Sources}

We rely on two primary sources to answer our RQs. We chose these sources for their focus, depth of analysis and information quality when discussing and resolving \hqc issues.

\subsubsection{Xanadu Discussion Forums (\xan)}
Our study primarily analyzed the \xan, an active discussion board hosted by Xanadu, a leading photonic \quantumcomputing firm \cite{xanaduforums, xanadubackground}. The board, featuring contributions from both community members and Xanadu staff, focuses on in-depth troubleshooting in quantum ML and \hqc algorithm development, using tools like the PennyLane library \cite{bergholm2018pennylane}. The technical discussions on \xan, especially those regarding bugs and other software issues, offer rich data for our analysis, a resource previously untapped by prior research focusing solely on quantum bugs. 

\subsubsection{Quantum Computing Stack Exchange (\stackexchange)}
The \stackexchange forum, akin to Stack Overflow for \quantumcomputing, serves as our study's secondary data source \cite{quantumstackexchange}. It features specialized discussions on \quantumcomputing challenges, including issues related to \hqc systems and applications. Its focused content, excluding basic programming queries, provided us with relevant data on complex \hqc issues that prior studies focused on bug identification have overlooked.

\subsubsection{Other Potential Data Sources}
Although we considered their inclusion, our study excluded GitHub repositories of \hqc applications as a data source due to prevalent data limitations. We identified only 540 \hqc application GitHub repositories (via GitHub keyword searching and inspection of GitHub's dependency graph), with over half of them being low-impact/educational projects or documentation examples. A significant portion (91\%) displayed minimal activity, evidenced by scant commits or issues, and lacked updates within the last year (as of Oct. 2023). Furthermore, rudimentary issue tracking and non-descriptive commit messages obstructed the extraction of informative bug-related information. 

\subsection{Data Collection}

\subsubsection{Search Terms}

We began the data collection process by deriving a set of search terms that would help us gather as many hybrid-specific threads as possible. The search terms were created and refined by two authors who both have at least two years of prior experience studying quantum computing and writing quantum applications in an academic setting. In developing the list of search terms, the authors did preliminary searches on the \xan and \stackexchange message boards for hybrid application-related posts to look for common terms that should be included as part of the list. Likewise, they looked at code for several well-known \hqc algorithms (\eg VQE and QAOA) to further refine the search term list and ensure appropriate coverage. Details of these search terms can be found in \Cref{tab:keywords}, with the complete list accessible in our replication package \cite{anonymous_repo}.
\looseness=-1

\begin{table}[]
\caption{Keywords used to identify relevant threads}
\label{tab:keywords}
\resizebox{\columnwidth}{!}{%
\begin{tabular}{l|l|l}
\textbf{Type} & \textbf{Rationale} & \textbf{Examples} \\ \hline
General & Capture spectrum of issues & ``bug'', ``invalid'' \\
QC-specific & Linked to QC processes & ``embed'', ``measure'' \\
PennyLane Template & Prominence of PennyLane in HQC & ``AmplitudeEmbedding'' \\
Hybrid Algorithm & Issues related to HQC algorithms & ``VQE'', ``QAOA''
\end{tabular}
}
\end{table}

\subsubsection{Raw Data Collection}

\begin{table}[]
\caption{Statistics about the discussion thread mining }
\label{tab:mining}
\resizebox{\columnwidth}{!}{%
\begin{tabular}{c|c|c|c}
\textbf{Forum} & \textbf{Mined Threads} & \textbf{Relevant Threads} & \textbf{Coded Threads} \\ \hline
{\rule{0pt}{1em}Xanadu} & 2,394 & 940 & 293 \\
Stack Exchange & 5,884 & 255 & 154 \\
\hline
\textbf{Total} & \textbf{8,278} & \textbf{1,215} & \textbf{447}
\end{tabular}%
}
\end{table}

The search term list guided our data collection from the \xan and \stackexchange using custom scripts written by the authors to automate the search for and collection of relevant \hqc issue discussions and posts, which helped streamline the process amidst the forums' vast data. These scripts curated URLs from the results, excluding duplicates, during searches conducted in Oct. 2023, with the archived threads detailed in our replication package~\cite{anonymous_repo}.  Notably, the range of relevant posts collected included posts from 2018 to late 2023, providing several years of data. 

Subsequent to URL curation, each thread was manually reviewed and scrutinized for relevance by one of the authors, prioritizing inclusivity to capture as many potential \hqc issue discussions as possible.  At this stage, due to the impracticality resulting from the volume and complexity of the data (particularly from \xan; see \Cref{tab:post_stats}), one author reviewed each thread to sift out false positives. Specifically, non-\hqc or non-bug threads were omitted. The manual review figures and exclusion of false positives are summarized in \Cref{tab:mining}, resulting in 940 relevant threads from \xan and 255 from \stackexchange retained for analysis.

\begin{table}[]\caption{Statistics about the coded discussion threads }
\label{tab:post_stats}
\resizebox{\columnwidth}{!}{%
\begin{tabular}{lc|c|c|c|c|c}
\multicolumn{2}{c|}{\textbf{Forum}}                                 & \textbf{\begin{tabular}[c]{@{}c@{}}Users\\ Involved\end{tabular}} & \textbf{\begin{tabular}[c]{@{}c@{}}Posts /\\ Comments\end{tabular}} & \textbf{\begin{tabular}[c]{@{}c@{}}Duration \\ (in days)\end{tabular}} & \textbf{\begin{tabular}[c]{@{}c@{}}Char \\ Count\end{tabular}} & \textbf{\begin{tabular}[c]{@{}c@{}}Word \\ Count\end{tabular}} \\ \hline
\multicolumn{1}{l|}{\multirow{3}{*}{\xan}}     & \textit{Min.} & 1                                                                 & 1                                                                   & 0                                                                      & 534                                                            & 77                                                             \\
\multicolumn{1}{l|}{}                               & \textit{Avg.} & 3                                                                 & 9.14                                                                & 68.52                                                                  & 9,619.78                                                        & 1,243.3                                                         \\
\multicolumn{1}{l|}{}                               & \textit{Max.} & 16                                                                & 86                                                                  & 1,239.53                                                                & 75,298                                                          & 8,196                                                           \\ \hline
\multicolumn{1}{l|}{\multirow{3}{*}{\stackexchange}} & \textit{Min.} & 1                                                                 & 1                                                                   & 0                                                                      & 272                                                            & 40                                                             \\
\multicolumn{1}{l|}{}                               & \textit{Avg.} & 3.11                                                              & 4.97                                                                & 95.61                                                                  & 4,406.82                                                        & 529.16                                                         \\
\multicolumn{1}{l|}{}                               & \textit{Max.} & 13                                                                & 20                                                                  & 1,311.54                                                                & 29,011                                                          & 2,089                                                          
\end{tabular}
}
\end{table}

\subsection{Coding of \hqc Discussion Threads}

In our coding framework, each identified issue in a thread was categorized into a top-level code. These top-level codes were developed by the authors based on their research, experience and from prior research of classical and quantum issue and bug categories \cite{li2022bug, catolino2019not}.  The categories were refined iteratively for alignment with the data collected from the threads. Sub-categories were delineated via \textit{open-coding} \cite{spencer2009card}:

\begin{itemize}

\item \textit{Independent Coding:} Two authors coded each identified issue in a thread.  This coding involved assigning a top-level code to each issue and labeling a sub-category that was open-coded by each author allowing for a more nuanced and detailed classification of each issue.  Sub-categories were later refined and, if necessary, condensed, merged and re-labeled by the two authors during regular reconciliation meetings.  This evolved into a shared code catalog.  This method, documented in our code catalog (see our replication package~\cite{anonymous_repo}), also involved extraction of relevant text from each thread, detailed issue descriptions, and if available, the case of the issue and how the issue was resolved and manifested.
\looseness=-1

\item \textit{Verification and Content Relevance:} The two coders ensured threads were relevant to \hqc and contained discussions of relevant issues and challenges faced by \hqc developers.

\item \textit{Reconciliation Meetings:} Authors resolved coding discrepancies through in-depth reconciliation meetings and discussions that generally lasted in the order of hours. Disagreements were rigorously discussed and resolved without the need to bring in a third author, though the option was available.  This step ensured consistency and reliability in the coding process mitigating potential bias. Since many of the threads contained multiple issues, the authors attempted to resolve discrepancies in coding through discussion, prioritizing coding consistency over inter-rate agreement metrics.

\end{itemize}

Our coding approach for threads was conducted in two distinct phases: a pilot coding of 20 \xan discussion threads followed by the coding of a representative sample of both \xan and \stackexchange threads (at 95\% confidence interval and 5\% error margin for each data source). In total, we coded 293 threads from \xan, including the pilot issues and a sample of 273, alongside 154 threads from \stackexchange.

From this analysis, we documented {377} issues on \xan and {154} issues on \stackexchange. Multiplicity of issues within single threads necessitated individual coding. Some threads were excluded from the count, such as those categorized as “Comprehension or Technical Questions.” These exclusions were adjudicated during reconciliation meetings that differentiated threads aimed at understanding \quantumcomputing or \hqc nuances from those reporting explicit issues. Overall, {531} real-world issues were coded, with {483} of these possessing identified causes within the domain of \hqc algorithms and applications.
\looseness=-1

\vspace{-.2cm}
\subsection{Analysis of Codes}
\label{sec:taxonomy}

\begin{figure*}[h]
\centering
\includegraphics[width=0.82\textwidth]{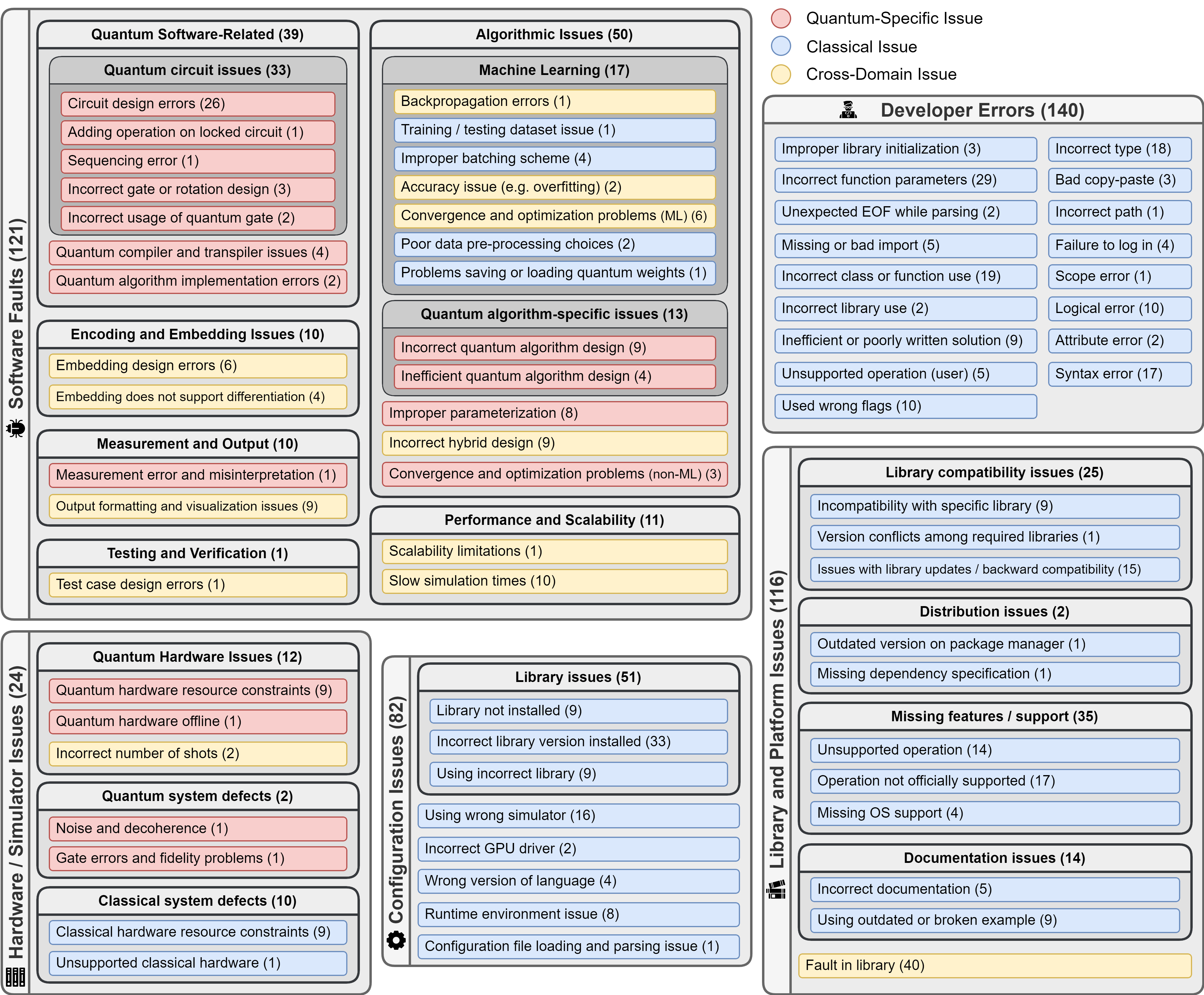}
\caption{Taxonomy of issues encountered by developers of \hqc applications} 
\label{fig:taxonomy}
\end{figure*}

After completing the coding process, the authors conducted an analysis to address RQ1-RQ5, which included:

\begin{itemize}
\item \textit{Developing a Taxonomy}: Addressing \textbf{RQ1}, this involved formulating a structured classification for common issues in \hqc applications. Originating from the coding phase, this taxonomy was refined through author discussions, resulting in a detailed four-level hierarchy presented in \Cref{fig:taxonomy}.% and discussed below.
    
\item \textit{Metrics Calculation and Analysis}: For \textbf{RQ2} to \textbf{RQ4}, the authors computed metrics from the coded data, to extract statistics with respect to domain classification, manifestations, and causes. Further, the taxonomy is color-coded to highlight distinctions between quantum-specific, classical, and cross-domain issues, with metrics offering a detailed overview of each issue’s frequencies and characteristics.
    
\item \textit{Identifying Recurring Challenges}: To answer \textbf{RQ5}, we analyzed recurring and predominant challenges faced by developers in the discussion threads, leading to the identification of common developer obstacles and the formulation of actionable recommendations.
\end{itemize}

\section{Results and Analysis}
\label{sec:results}

\subsection{RQ1: Recurrent Issues in \hqc Applications}

Our study is primarily aimed at identifying the spectrum of issues that \hqc developers grapple with when building hybrid applications, extending beyond the realm of conventional `bugs' that prior research has focused on. As we sifted through the forum discussions, it became evident that developers were wrestling with challenges more complex than code faults, but rather broader issues (related to hardware/simulators, library usage and configuration, \etc) that significantly hampered \hqc application development. Many of these difficulties are symptomatic of the infancy of \hqc computing and the brisk evolution of its platforms and practices in QC. Our taxonomy thus captures a wider array of obstacles, including recurrent bug patterns, extracted from the discourse in issue threads. 

After categorizing {531} real-world issues from the \xan and \stackexchange, we analyzed the results and developed a taxonomy as seen in \Cref{fig:taxonomy}. The taxonomy is organized hierarchically, beginning with five high-level classifications that represent the broad areas where issues arise in \hqc applications: 

% \begin{itemize}

\textbf{Software Faults (121)} are issues traditionally referred to as ``bugs''.  They are faults inherent to the software of \hqc applications, where the code or its design is flawed, leading to  failures or incorrect program execution. For instance, a software fault includes a circuit design error where the developer \rev{uses incorrect gate operations or mismanages qubits}.~\cite{xan_poor_circuit_design}.

\textbf{Hardware/Simulator Issues (24)} concern problems related to the physical quantum hardware or quantum computer simulators that \hqc applications run on. Issues can range from constraints within the classical simulation environment (\eg limitations on the number of qubits permitted by a simulator~\cite{xan_classical_constraints}) to quantum hardware resource limitations (\eg delayed executed due to long queue length on IBM quantum hardware~\cite{xan_stuck_in_queue}) and even quantum hardware being offline~\cite{xan_hardware_offline}. 

\textbf{Configuration Issues (82)} arise from incorrect or suboptimal setup of the computing environment and/or dependencies used by the developer in building the hybrid application. Unlike the category “Library and Platform Issues” (described below), the problems in this category are generally caused by the developer that uses quantum libraries and can be resolved within the application using such libraries. Examples include having the incorrect version of a library or dependency installed~\cite{xan_wrong_library_version}, using a wrong simulator \cite{xan_wrong_simulator}, and errors in setting up a runtime environment \cite{se_runtime_problem}.

\textbf{Library and Platform Issues (116)} faults and other defects within the libraries and platforms used to develop \hqc applications.  The issues are generally out of the control of the application developer, who must rely on the platform or library developers to resolve them. Issues in this category include backwards compatibility after library updates \cite{xan_backwards}, operations that are not officially supported by a library \cite{xan_unsupported_operation}, incorrect documentation \cite{xan_incorrect_documentation} and bugs in the platforms themselves \cite{xan_bug_in_library}.

\textbf{Developer Errors (140)} are mistakes made by application developers, often involving incorrect code syntax \cite{xan_basic_python}, incorrect types\cite{xan_incorrect_type}, logic errors \cite{xan_logic_error}, or using incorrect function parameters \cite{xan_incorrect_parameter}. They are the direct result of human error during the SE process and are generally fixable by correcting the application's code. 
Moreover, these issues almost exclusively fall in the realm of classical issues.  These types of issues are well-studied in other research, but have been included in the taxonomy for completeness.

The \hqc issues categorized according to our taxonomy amount to 483 issues in total, described in 447 discussion threads. In an additional 48 discussion threads, we encountered issues which we could not categorize into the taxonomy due to limited and/or unclear information in the forum thread.  Common examples of these threads would be where an original post would go unanswered or where the original poster failed to follow-up to responses in the thread \cite{se_unknown, xan_unknown}.

Each top-level category is further divided into sub-categories, providing more granular insight into specific issues that were observed. To illustrate the frequency of each encountered issue, the taxonomy in \Cref{fig:taxonomy} also provides counts for each observed issue.  This hierarchical structure enables a systematic examination of recurring issues in \hqc application, from general phenomena to specific instances, covering a variety of possible problems a \hqc developer could encounter. 

Note that our taxonomy includes most bugs and issues outlined in previous works \cite{paltenghi2022bugs, luo2022comprehensive, zhao2021bugs4q, li2022bug}, but at different granularities.  For instance, Configuration Issues in our taxonomy are broken down into several sub-categories providing more detail as to the issue encountered.  This contrasts with prior research that lumped all configuration bugs into a single "misconfiguration" category \cite{paltenghi2022bugs, luo2022comprehensive, zhao2023empirical}.

\rev{Some issue categories, like Circuit Design Errors, are admittedly somewhat broad. The prevalence of these errors aligns with prior studies \cite{paltenghi2022bugs, li2022bug, luo2022comprehensive, zhao2023empirical, zhao2021bugs4q}. However, our dataset revealed unique, non-repeating errors (\eg no measurement, multiple measurements, wrong gate, overlapping gates, wrong qubit order), making finer granularity difficult. This variability in issues encountered underscores the complexities of quantum programming and suggests further research focused specifically on Circuit Design Errors could create more specific subcategories for better classification.}

\rev{Consequently, there are a handful of bug types, mostly relating to algorithmic and circuit design issues, found in prior works that are not explicitly mentioned in our taxonomy~\cite{paltenghi2022bugs, luo2022comprehensive}.  These bug types are more granular sub-categories of issues, like the aforementioned Circuit Design Errors, found in our taxonomy.  A complete chart comparing the issue categories found in taxonomy compared to bugs found in prior works in contained in our replication package~\cite{anonymous_repo}.}

Importantly, though, the taxonomy presented in this paper captures many more bugs and issues that are not contained in prior works due to our focus on \hqc applications.  These include issues such as software faults related to parameterization, convergence and optimization, encoding and decoding, machine learning-specific faults, hardware and simulator issues and, of course, issues specific to \hqc application design (see \Cref{fig:taxonomy}). \rev{This is also the case for classical developer errors.  Prior works have typically only identified a few of these faults, such syntax errors and type errors \cite{paltenghi2022bugs, luo2022comprehensive, zhao2023empirical}. In this study, we were able to capture many more types of developer errors, such as missing or bad imports, used wrong flags, improper library initialization, incorrect function parameters and many others.}
\looseness=-1

\rev{Our taxonomy is not just a high-level classification of issues in \hqc applications, though. It also serves as a practical tool for debugging and testing. Each category and sub-category offers a structured approach to identifying and addressing specific issues that \hqc developers encounter. For instance, developers facing a fault or an error in a library or platform can use the taxonomy's Library or Platform Issue category to help determine if the problem is due to missing features, poor documentation, version compatibility, or other specific bugs. This detailed guidance helps developers systematically troubleshoot, diagnose, and fix issues, potentially enhancing the efficiency and effectiveness of their debugging processes.}

\rev{Moreover, the counts in the taxonomy provide insights into the prevalence of different issues in \hqc applications. Developers and testers can use this data to prioritize their efforts based on the most common and critical issues. For instance, our taxonomy indicates that ``Configuration Issues'' are a particularly common occurrence. As a result, developers can focus on improving their setup and configuration processes, while testers can design specific test cases to catch these issues early. These metrics help identify recurring patterns, enabling developers to take proactive measures to prevent common issues in quantum software.}

We further discuss a number of issue categories and sub-categories, with examples, in the remaining RQs. Definitions for all categories can be found in our replication package \cite{anonymous_repo}.

\subsection{RQ2: Origins of Issues in \hqc Applications}

Our taxonomy classifies issues by their origins in \hqc systems, where the origins were identified based on whether the fix to the issue was achieved in the quantum and/or classical subsystems of an \hqc application. These classifications are delineated with color codes for clarity on their distribution in \Cref{fig:taxonomy}. 
We group the identified \hqc issues into three categories:

\textbf{Quantum-Specific Issues (76)} arise solely within the quantum  component of an \hqc application, including quantum circuit design \cite{xan_poor_circuit_design, xan_gate_operation}, gate and rotation operations~\cite{se_gate_operation} and quantum algorithm implementation \cite{xan_algo}.
\looseness=-1

\textbf{Classical Issues (316)} are found in the conventional or classical computing portions of an \hqc application \cite{xan_basic_python, xan_backwards, xan_wrong_library_version}. They can also include some machine learning issues as well as classical hardware defects \cite{xan_bingo}.

\begin{figure}[t]
\centering
\includegraphics[width=.8\linewidth]{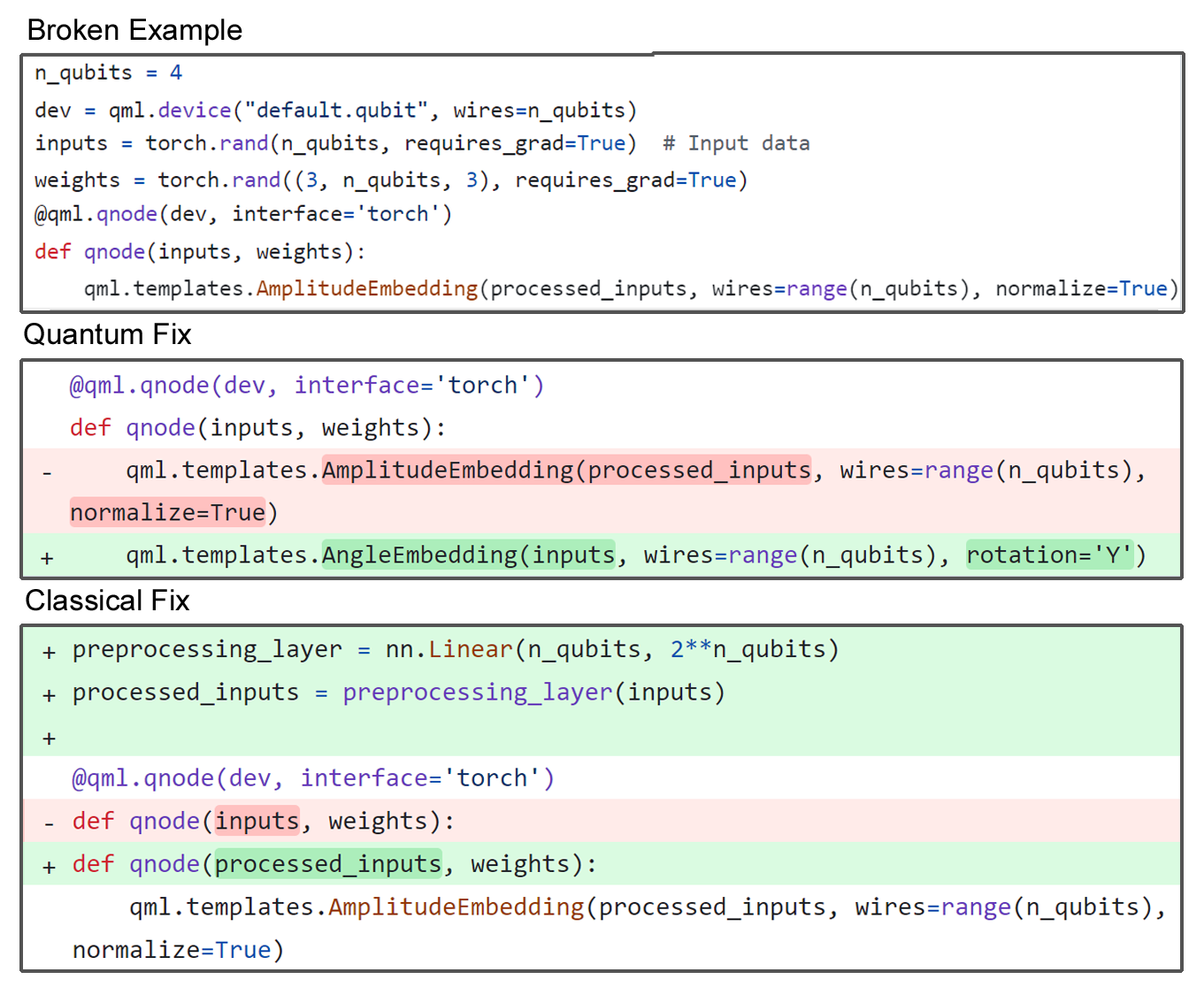}
\caption{Platform Quirk Example with Cross-Domain Fixes}
{\footnotesize 
AmplitudeEmbedding in PennyLane is not differentiable.}
\label{fig:cross}
\end{figure}

\textbf{Cross-Domain Issues (91)} manifest at the nexus of quantum and classical realms, often being addressed through adjustments in either component. More specifically, these issues are distinguished by their ability to manifest in either the quantum or classical parts of the system and can typically be resolved through modifications to either subsystem.  For example, as shown in \Cref{fig:cross}, an embedding fault is a cross-domain issue because it can be resolved by changing the quantum embedding circuit(s) or by adapting the classical algorithm and output to conform with the embedding scheme's constraints. Other issues include incorrect hybrid algorithm design \cite{xan_ml} and machine learning-related issues \cite{xan_hqc_design}. Notably, this category has not previously been identified in prior quantum bug studies~\cite{paltenghi2022bugs, luo2022comprehensive}.

% \end{itemize}

\Cref{tab:man-by-orig} shows the distribution of issues across the three categories. The predominance of classical issues implies the continued relevance and necessity of traditional debugging, error detection, and fault resolution techniques in \hqc applications. Moreover, the notable share of cross-domain bugs points to unique challenges that blend quantum and classical computing, as these issues may require a nuanced or more specialized understanding of both quantum and classical computing principles for effective resolution (see \Cref{fig:cross}). Finally, although quantum-specific issues were less frequent than the other issues, they nonetheless highlight the need for developers to possess special insights into quantum computing principles and employ targeted troubleshooting methods~\cite{paltenghi2022bugs}.

\subsection{RQ3: Manifestation of \hqc Issues}

\begin{table}[]
\caption{Instsances of Issue Manifestations by Origin}
\label{tab:man-by-orig}
\resizebox{\columnwidth}{!}{%
\begin{tabular}{l|ccc|c}
\textbf{Manifestation} & \textbf{Quantum-Specific} & \textbf{Classical} & \textbf{Cross-Domain} & \textbf{All Issues} \\ \hline
\textbf{Crash} & 51 (67.11\%) & 257 (81.59\%) & 51 (55.43\%) & 359 (74.33\%) \\
\textbf{Incorrect Output} & 15 (19.74\%) & 28 (8.89\%) & 26 (28.26\%) & 69 (14.29\%) \\
\textbf{Slow Execution} & 6 (7.89\%) & 13 (4.13\%) & 11 (11.96\%) & 30 (6.21\%) \\
\textbf{Unknown} & 3 (3.95\%) & 14 (4.44\%) & 2 (2.17\%) & 19 (3.93\%) \\
\textbf{Warning} & 1 (1.32\%) & 3 (0.9\%) & 2 (2.17\%) & 6 (1.24\%) \\
\hline
\textbf{Total} & 76 (100\%) & 315 (100\%) & 92 (100\%) & 483 (100\%)
\end{tabular}%
}
\end{table}

To answer RQ3, we performed an analysis of how issues encountered in the development of \hqc applications manifest themselves in order to understand how to mitigate them. We found that \hqc application issues manifested as follows:

\textbf{Crash (359):} This is the most direct manifestation of \hqc issues, where the application fails to execute to completion, typically resulting in an error message. This abrupt termination %of the program
could stem from a variety of issues, including quantum circuit and logic errors, classical code exceptions, or incompatibilities between quantum and classical components \cite{xan_incorrect_type, xan_hqc_design}.

\textbf{Incorrect Output (69):}  At times, the application or algorithm may complete its execution but yield an output that deviates from the expected result \cite{xan_algo}.  For instance, this discrepancy could be a consequence of logical errors in algorithm implementation, erroneous quantum gate operations, or faulty data processing on the classical side.

\textbf{Slow Execution (30):} The performance of \hqc applications is often a balancing act between the theoretical speedup offered by \quantumcomputing and the practical limitations of current technology. Slow execution typically arises from inefficient algorithm or circuit designs, limited availability of or long queues for quantum hardware resources, suboptimal simulator performance, or even classical hardware constraints \cite{xan_slow_exec}.

\begin{figure}
\centering
\includegraphics[width=.8\linewidth]{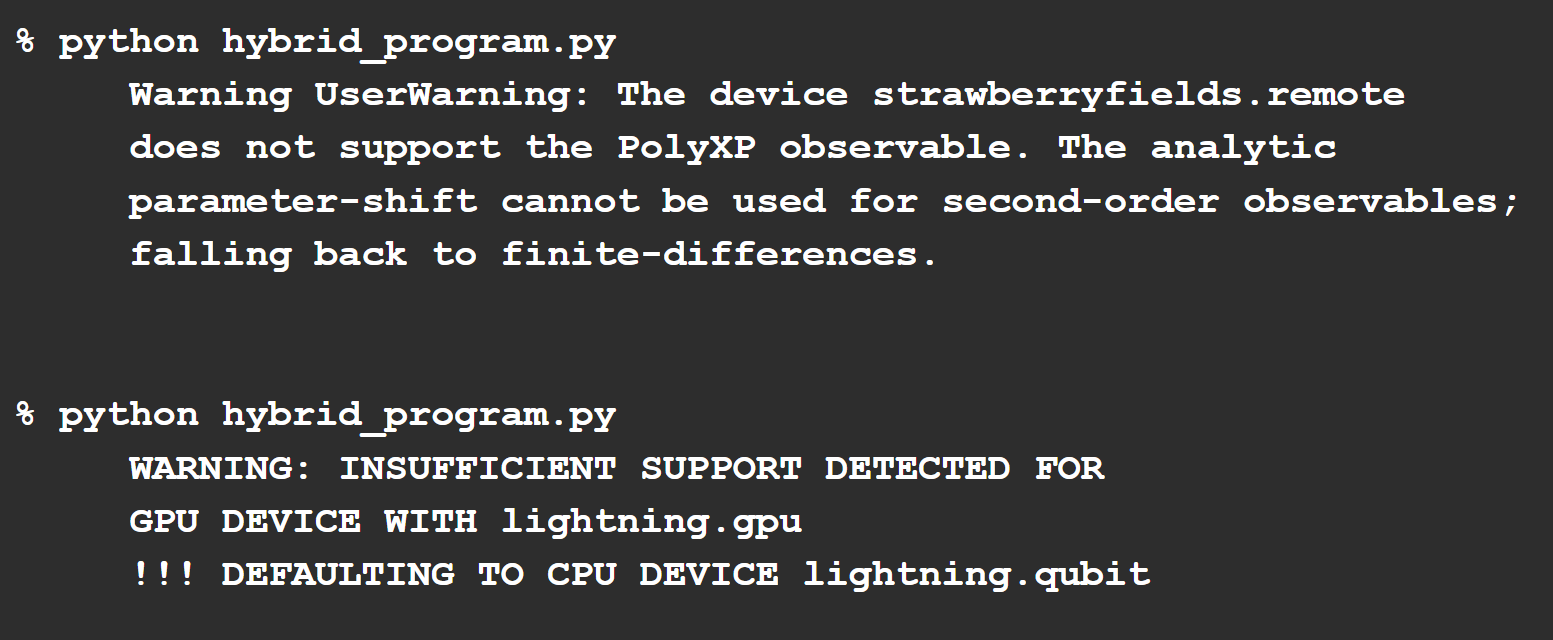}
\caption{Example PennyLane Warning Messages}
{\footnotesize Two warning messages encountered by devs in PennyLane that change application behavior regardless of what the was specified in the code~\cite{xan_warning_2}.}
\label{fig:warning}
\vspace{-.25cm}
\end{figure}

\textbf{Warnings (19):}  They serve as a preemptive indicator of potential issues in \hqc applications. As shown in \Cref{fig:warning}, they may inform the user of a suboptimal operation or an automatic fallback to a different simulator or function due to some detected incompatibility or inefficiency \cite{xan_warning, xan_warning_2}.

\textbf{Unknown (6):}  Some analyzed threads did not provide enough information to ascertain how an issue manifests.  We labeled these issues as having an unknown manifestation~\cite{xan_unknown_manifestation}.

Notably, we did not encounter issues like corrupted data or hanging applications that are common in classical software~\cite{tan2014bug}. The absence of corrupted data issues likely stems from limited file processing in current \quantumcomputing and \hqc applications. While hanging is theoretically possible, it was not observed, possibly due to the stateful nature of quantum components and built-in error handling in platforms like Qiskit and PennyLane, which tend to force crashes. 

\Cref{tab:man-by-orig} provides an overview of issue manifestations across all identified issues encountered by developers in the threads %(including those with an unknown cause). 
These figures contrast starkly with those from classical software studies, where incorrect output predominates \cite{cotroneo2016bugs, tan2014bug, di2017comprehensive}. The overall higher incidence of crashes in \hqc applications likely reflects the relative immaturity and lack of robust error handling in quantum platforms and libraries.  The rapidly changing nature of quantum software development tools, coupled with the intricate challenges posed by integrating quantum components, no doubt leads to an increased propensity for crashes.  Additionally, the steep learning curve in developing applications leading to algorithmic errors within quantum components may also contribute to the prevalence of crashes. 

Quantum-specific issues tended to have more instances of incorrect output and fewer crashes compared to purely classical issues. This may be due to the inherent probabilistic nature of quantum circuit outputs, where even a correctly functioning quantum algorithm might yield varying or unexpected results due to quantum phenomena. This could also result from incorrect quantum logic or poor circuit design, again underscoring the importance of developer education.  This is in line with prior work focusing on bugs in quantum platforms \cite{paltenghi2022bugs}.

Classical issues appeared to have high instances of crashing. The reasons for this could be manifold. The Library and Platform Issues category from the taxonomy account for 24\% of all issues sampled suggesting a landscape where the tools and frameworks like Qiskit and PennyLane are still in a state of flux with updates and changes potentially introducing new points of failure.  Furthermore, the quantum learning curve is very steep and can lead to a higher incidence of classical issues as evidenced by the high number of developer errors.

Cross-domain issues were still crash dominant but had higher instances of incorrect output and slow execution compared to quantum-specific and classical issues. This likely results from our classifications, where slow simulation times may stem from classical hardware issues or quantum algorithm design flaws. Also, cross-domain issues, like embedding issues and hybrid design errors, emerge at the intersection of quantum and classical computing, where mismatches in data formats, incompatible logic flows, or synchronization problems between the two domains can lead to more diverse and complex manifestations of bugs and issues.

\looseness=-1

\begin{table}[]
\caption{\rev{Instances of Manifestations by Issue Type}}
\label{tab:manifestations_by_category}
\resizebox{\columnwidth}{!}{%
\begin{tabular}{l|ccccc}
\textbf{Issue Type}  & \textbf{Crash} & \textbf{\begin{tabular}[c]{@{}c@{}}Incorrect \\ Output\end{tabular}} & \textbf{\begin{tabular}[c]{@{}c@{}}Slow \\ Execution\end{tabular}} & \textbf{Unknown} & \textbf{Warning} \\ \hline
\textbf{Algorithmic} & 29 & 19 & 1 & 1 & 0 \\
\textbf{Config} & 68 & 4 & 7 & 1 & 2 \\
\textbf{Developer Errors} & 121 & 14 & 3 & 0 & 2 \\
\textbf{Quantum Specific} & 31 & 4 & 0 & 3 & 1 \\
\textbf{Encoding / Embedding} & 8 & 1 & 0 & 1 & 0 \\
\textbf{Library / Platform} & 85 & 17 & 2 & 10 & 2 \\
\textbf{Measurement / Output} & 3 & 7 & 0 & 0 & 0 \\
\textbf{Performance / Scalability} & 1 & 0 & 10 & 0 & 0 \\
\textbf{Hardware / Simulator} & 13 & 2 & 8 & 0 & 1 \\
\textbf{Testing / Verification} & 0 & 1 & 0 & 0 & 0 \\
\textbf{Unknown Cause} & 34 & 10 & 2 & 1 & 1 \\
\hline
\textbf{Total} & 393 & 79 & 33 & 17 & 9
\end{tabular}%
}
\end{table}

\rev{When examining issue manifestations in the taxonomy (see \Cref{tab:manifestations_by_category}), most issue types predominantly cause crashes. However, Algorithmic Issues are more likely to produce higher instances of Incorrect Output, comparatively. This is due to the nature of quantum algorithms, where slight deviations in quantum logic, gate operations, or noise can result in significant discrepancies in output. Unlike crashes, these subtle errors in algorithmic processes may not trigger immediate failures, leading instead to incorrect results that are harder to detect.}

\rev{Configuration Issues had higher instances of manifesting with slower execution, likely due to suboptimal setup of computing environments or dependencies, which can hinder performance and efficiency. Library and Platform Issues showed notable instances of Incorrect Output and Unknown Manifestations. Incorrect output often stemmed from bugs or undocumented behaviors in libraries, while manifestations with an unknown cause were due to users lacking sufficient data or error messages from platforms for accurate diagnosis. Performance and Scalability Issues, along with Hardware/Simulator Issues, had higher instances of slower execution. This is likely due to the probabilistic nature of quantum hardware, unoptimized simulators, and limited hardware resources.}

\subsection{RQ4: Causes of \hqc Issues}

Creating \hqc applications introduces unique challenges where the complexity of quantum mechanics meets traditional computing. In exploring the specific root causes of issues encountered by developers in \hqc applications, our study identified the following causes of issues:

\textbf{Programmer Errors (338)}: Discrepancies introduced by developers, ranging from simple syntactical mistakes to deeper conceptual misunderstandings of quantum operations or algorithms, resulting in incorrect or inefficient code.

\textbf{Platform Issues (130)}: Stem from inconsistencies, limitations, or defects within the development environments and libraries themselves used by \hqc developers.  They manifest in a variety of ways within the \quantumcomputing libraries, such as backward compatibility challenges when library versions change or unexpected behaviors due to poor documentation.

\begin{figure}[t]
\centering
\includegraphics[width=.8\linewidth]{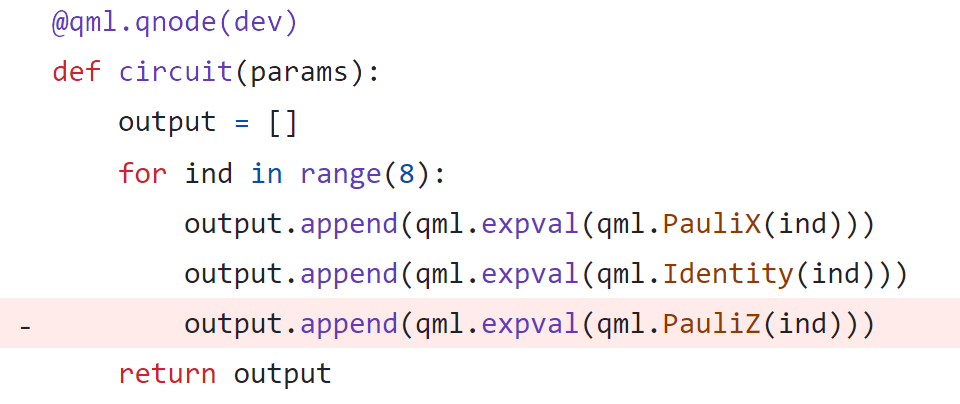}
\caption{Platform Limitation Example}
{\footnotesize Attempting multiple measurements on the same wire with non-commutable observables (PauliX / PauliZ) causes errors due to PennyLane restrictions~\cite{xan_champion}}

\label{fig:limitation}
\end{figure}

\textbf{Platform Limitations (3)}: Refer to the inherent constraints set by the development platforms and libraries, which can restrict the capabilities available to developers.  Unlike \textit{Platform Issues}, which are unintended faults in the system, platform limitations are known restrictions, such as computational or resource limitations, that developers must navigate within. See Fig. \ref{fig:limitation} for an example.

\textbf{Hardware Issues (12):} Problems that originate from the \quantumcomputing or classical hardware running quantum simulators. These include the physical limitations of quantum devices, such as decoherence, error rates, or connectivity constraints between qubits. While hardware issues are less prevalent in SE stages that primarily utilize simulators, they become significant when applications are deployed on actual quantum hardware where physical phenomena can introduce errors.

\begin{table}[]
\begin{center}
\caption{Instances of Issue Causes by Origin}
\label{tab:bug-causes-by-origin}
\resizebox{\columnwidth}{!}{%
\begin{tabular}{l|cccc}
\multicolumn{1}{c|}{\textbf{}} & \textbf{Quantum-Specific} & \textbf{Classical} & \textbf{Cross-Domain} & \textbf{All Issues} \\ \hline
\textbf{Programmer Error} & 63 (82.89\%) & 236 (74.92\%) & 39 (42.39\%) & 338 (69.98\%) \\
\textbf{Platform Issue} & 10 (13.16\%) & 67 (21.27\%) & 53 (57.61\%) & 130 (26.92\%) \\
\textbf{Platform Limitation} & 0 & 3 (0.95\%) & 0 & 3 (0.62\%) \\
\textbf{Hardware Issue} & 3 (3.95\%) & 9 (2.86\%) & 0 & 12 (2.48\%) \\
\hline
\textbf{Total} & 76 (100\%) & 315 (100\%) & 92 (100\%) & 483 (100\%)
\end{tabular}%
}
\end{center}
\end{table}

Table \ref{tab:bug-causes-by-origin} shows statistics for the causes of 483 categorized issues. Programmer Error is the predominant cause across all domains, highlighting the complexity of \quantumcomputing concepts for developers and their application in \hqc contexts. This high rate of programmer-induced issues suggests the need for better educational resources, effective debugging tools, and improved guidance on \hqc software engineering practices.

The significant portion of platform-induced issues highlights the current state of quantum SE tools, which are still in active development and refinement. As can be seen in \Cref{fig:taxonomy}, a large number of encountered issues (116) are attributable to Platform and Library Issues.  As these platforms mature and stabilize, we might expect a reduction in platform-related issues.  \rev{In the meantime, platform developers should work more closely with application developers to reduce these issues.}

It is notable that hardware-induced issues made up a small portion of our dataset, suggesting that while quantum hardware is a concern, the current focus of running \hqc applications on simulators during development and this NISQ-era of \quantumcomputing guards applications from many quantum hardware-specific issues, particularly those intrinsic to quantum physics.

\Cref{tab:bug-causes-by-origin} shows a balanced distribution between platform-induced issues and programmer errors in cross-domain contexts. This reflects the complexity of integrating quantum and classical systems, where issues can often be resolved with either quantum or classical fixes. It appears that both developers and platforms are still refining integration methods as \hqc applications become more prevalent.

\rev{Additionally, when examining the causes of errors across the taxonomy categories, the results appear to be relatively straightforward. Programmer Errors caused all Developer Errors in the taxonomy and over 90\% of Algorithmic, Configuration, Encoding and Embedding, and Quantum Software-Related Issues. Platform-induced errors caused nearly all Library or Platform Issues. Hardware/Simulator Issues were roughly equally caused by Platform Issues and Hardware Issues, reflecting a dual reliance on both hardware and the software platforms managing an \hqc application's operation.}

\rev{Lastly, \Cref{tab:manifestation_cause} provides some insight on the relationship between manifestations of issues and their causes. 
As discussed above, crashes are the predominant manifestation, which are primarily attributed to Programmer Errors. This indicates that a substantial proportion of stability problems in \hqc applications falls on developers. Additionally, with nearly a quarter of crashes attributable to Platform Issues, quantum computing platforms themselves, such as Qiskit and PennyLane, should continue to focus on refinement and robust error-handling mechanisms to enhance their. This is particularly true in light of the small percentage of hardware issues causing crashes.}

\rev{Similarly, Programmer Error accounted for nearly two-thirds of issues resulting in Incorrect Output.  Meanwhile, Platform Issues accounted for over a quarter. This again highlights the dual need for improved developer training and support as well as efforts by platforms to improve stability.}

\rev{Issues causing Slow Execution were mainly due to Platform Issues and Programmer Errors, suggesting that developers should continue to try to optimize their quantum algorithms and platforms should focus on enhancing performance. Although less frequent, issues resulting in Warnings highlight the need for better diagnostic and debugging tools, like linters, to preemptively address potential problems before they escalate.}

\begin{table}[h]
\caption{\rev{Instances of Manifestation Type Based on Cause}}
\centering
\resizebox{\columnwidth}{!}{%
\begin{tabular}{l|ccccc}
\textbf{} & \textbf{Crash} & \textbf{\begin{tabular}[c]{@{}c@{}}Incorrect \\ Output\end{tabular}} & \textbf{\begin{tabular}[c]{@{}c@{}}Slow \\ Execution\end{tabular}} & \textbf{Warning} & \textbf{Unknown} \\ \hline
\textbf{Platform Issue} & 82 (20.9\%) & 21 (26.6\%) & 18 (54.5\%) & 2 (28.6\%) & 7 (36.8\%) \\
\textbf{Programmer Error} & 268 (68.2\%) & 48 (60.8\%) & 10 (30.3\%) & 3 (42.9\%) & 9 (47.4\%) \\
\textbf{Hardware Issue} & 8 (3.5\%) & 0 & 3 (9.1\%) & 1 (14.3\%) & 0 \\
\textbf{Platform Limitation} & 1 (2.0\%) & 0 & 0 & 0 & 2 (10.5\%) \\
\textbf{Unknown Cause} & 34 (8.7\%) & 10 (12.6\%) & 2 (6.1\%) & 1 (14.3\%) & 1 (5.3\%) \\ \hline
\textbf{Total} & 393 & 79 & 33 & 7 & 19
\end{tabular}%
}
\label{tab:manifestation_cause}
\end{table}

\vspace{-.2cm}

\subsection{RQ5: Challenges Resolving \hqc Bugs}
\label{sec:RQ5}

In the course of our study, we observed several predominant and recurring challenges that developers of \hqc applications faced. We further analyze these challenges below: 

\textbf{Integration of New Features:} A notable hurdle, representing 6.4\% of encountered issues, was the lack of support for certain quantum operations, demanding workarounds. Open-source platforms like PennyLane responded swiftly, often incorporating these operations in subsequent versions, and sometimes requesting bug reporters to contribute enhancements (see this thread as an example: \cite{xan_champion}). Proactive feature expansion by libraries, mirroring the functionality of quantum hardware in simulators, is recommended for better user support.

\textbf{Keeping Pace with Platform Evolution:} The rapid evolution of quantum platforms is a double-edged sword. On one hand, it assists developers by increasing functionality; on the other, it can be prone to introducing errors and faults. As the taxonomy shows, platform and library compatibility issues accounted for 5.2\% of the issues encountered. We observed 15 instances where platform updates caused backward compatibility issues with other libraries used by developers. Additionally, the deprecation of entire sub-modules like Qiskit Aqua and Ignis~\cite{qiskitdeprecation} has rendered code, documentation, and resources obsolete. This is reflected in the fact that 1.9\% of all issues were caused by broken examples. Moreover, 6.8\% of developer issues arose from library-related bugs and issues, primarily due to outdated or incorrect library versions or dependency conflicts.  Such rapid changes can confuse developers and disrupt the development of \hqc applications. \rev{These issues can easily be prevented with increased testing by platforms and better communication between platform and application developers.}
\looseness=-1

\begin{figure}[t]
\centering
\includegraphics[width=.8\linewidth]{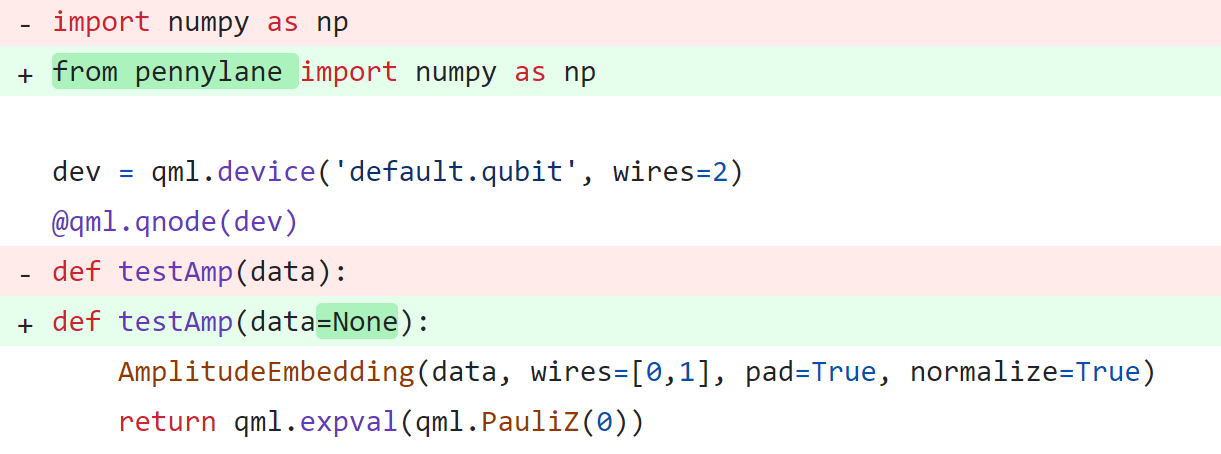}
\caption{Platform Quirks Examples}
{\footnotesize Redefining testAmp() to take a keyword argument resolves the error since AmplitudeEmbedding only accepts non-differentiable data.}
\label{fig:quirks}
\end{figure}

\textbf{Platform Quirks:} Libraries' peculiarities, such as PennyLane's custom NumPy library or the requirement that QNode arguments be non-differentiable if passed as keywords, can cause non-intuitive issues (see \Cref{fig:quirks}) and highlight the practical challenges \hqc developers face. \rev{In many cases, documentation failed to identify these quirks, and developers were unaware of them until they were raised in forums. To mitigate this, we suggest that platform developers provide clear documentation on quirks, compatibility guides, and tools to switch between library versions. Additionally, implementing warning or descriptive runtime error logs can help developers identify and resolve these issues more efficiently.}

\textbf{The Quantum Learning Gap:} Around 10.4\% of issues stemmed from Algorithm Issues in \hqc applications. Another 10.1\% of issues were Quantum Software-Related or quantum Encoding and Embedding Issues.  These issues most likely stemmed from a lack of understanding of quantum mechanics. Yet, a significant portion of other issues encountered, such as those in the Configuration Issues and Developer Errors categories, stem from classical errors rooted in the misunderstanding of the requirements of quantum platforms.  This underscores the necessity for developers to have a robust grasp of both quantum and classical computing principles when developing \hqc applications.

\rev{For application developers, we recommend developing formal debugging and testing procedures tailored to \hqc applications to address these issues, particularly algorithmic bugs. Ensuring environment consistency, such as using containerization technologies or virtual environments, can prevent many configuration-related issues. Regularly updating libraries and dependencies helps avoid compatibility problems and benefits from the latest improvements and bug fixes. However, care must be taken to manage library updates to avoid backward compatibility issues, such as by testing updates in a controlled environment before full deployment.} 

\rev{Platforms and frameworks can reduce the learning gap by maintaining up-to-date documentation and providing clear guidance on common pitfalls.
Additionally, continuous learning through courses, workshops, and community engagement is essential to keep up with the rapidly evolving field of quantum computing. Engaging with other quantum developers through forums, conferences, and collaborative projects can provide support and facilitate the sharing of best practices.}

\textbf{Expanding Developer Resources:} While resources such as the PennyLane and Qiskit documentation provide considerable support to developers of \hqc applications, there were still 48 issues (9\%) that remained unresolved in forums.  Some of these resulted from poor issue reporting or a failure to follow up with requested information (which is also a significant issue in purely classical systems \cite{chaparro2019assessing}).  This indicates a need for more extensive education and training in quantum development.  Moreover, there is concern that a significant amount of quantum knowledge could become sequestered within proprietary industry and government projects, limiting widespread learning and development.

\rev{We recommend quantum platforms implement streamlined error reporting mechanisms, like the Xanadu Discussion Forums, that allow application developers to quickly and efficiently identify and resolve issues. Also, establishing feature request platforms can foster better communication between platform developers and applications developers, ensuring that the most pressing needs are addressed quickly and efficiently.}

In sum, resolving the challenges and issues that arise during the development of \hqc applications is multi-faceted. By addressing these issues through improved development practices, comprehensive and current documentation, and robust platform designs that account for backward and cross-library compatibility, developers can build more reliable hybrid applications and engage in more streamlined resolution processes when an issue is encountered.

\section{Threats to Validity}

\textbf{Internal Validity}.
Our study may be influenced by the choice of data sources, \xan and \stackexchange, possibly omitting other \hqc issues. Single-author thread removal and manual coding introduced subjectivity, despite spot-checks and reconciliation efforts to ensure consistency. The manual nature of the coding could inherently allow for human error in issue identification and classification. We attempted to mitigate some of these threats by applying a rigorous coding methodology.

\textbf{External Validity}.
The generalizability of our findings is restricted by the selected forums, which might not be fully generalizable to all \hqc applications. The rapid evolution of \quantumcomputing could limit the long-term relevance of our issue taxonomy.

\textbf{Construct Validity}.
The nascent quantum software field's fluid definitions may affect the construct validity. Our taxonomy, reflecting the current state of \hqc understanding, might require updates as the industry matures and evolves.

\textbf{Replicability}.
While detailed methods and documentation aim to enhance replicability, the fast-paced changes in \quantumcomputing and inherent subjectivity in coding could lead to different results in future replications of this study.

\section{Conclusions}

We examined, analyzed, and categorized 531 real-world issues, discussed in quantum-focused forums, to construct a comprehensive taxonomy of issues encountered by developers working on \hqc applications. The taxonomy identifies key areas where developers struggle, organizing them into a clear framework that can assist developers in identifying and addressing common problems more efficiently.  We discussed these recurring challenges and provided actionable recommendations to address them. Lastly, we analyzed how \hqc bugs manifest, their causes, and which domain they predominated.

Looking ahead, \hqc application development is still in its infancy and there is a clear need for SE perspectives.  This includes designing and improving tools for debugging, enhancing learning materials for developers, and ensuring that the platforms and libraries used continue to incorporate new features while maintaining compatibility. Addressing these needs will help make working with \hqc systems more accessible, reliable, and efficient, paving the way for broader adoption and optimization of \hqc systems.

\bibliographystyle{IEEEtran}
\balance
\bibliography{references}

\end{document}